\documentclass[prc,showpacs,preprint]{revtex4}
\usepackage{amssymb}

\usepackage{graphicx}
\usepackage{dcolumn}
\usepackage{bm}
\usepackage{multirow}

\begin{document}

\preprint{1.0}
\title{Thermal shape fluctuation model study of  the 
giant dipole resonance  in $^{152}$Gd }
\author{A. K. Rhine Kumar}
\email{rhinekumar@gmail.com}
\author{P. Arumugam}
\affiliation{Department of Physics, Indian Institute of Technology Roorkee,
Uttarakhand - 247 667, India}
\date{\today}

\begin{abstract}
We have studied the giant dipole resonance (GDR) in the hot and rotating
nucleus $^{152}$Gd within the framework of thermal shape fluctuation model
(TSFM) built on the microscopic-macroscopic calculations of the free energies
with a macroscopic approach for the GDR. Our results for GDR cross sections are in good agreement with the experimental values except for a component
peaking around 17 MeV where the data has large uncertainties. Such a component
is beyond our description which properly takes care of the splitting of GDR components due to the deformation and Coriolis effects. Around this 17 MeV lies the half maximum in experimental cross sections, and
hence the extracted GDR widths and deformations  (estimated from these widths) turn out to be overestimated and less reliable. Reproducing these widths with empirical formulae could conceal the information
contained in the cross sections. Fully microscopic GDR calculations
and a more careful look at the data could be useful to understand the GDR component around 17 MeV. We also discuss the occurrence of $\gamma$-softness in the free energy surfaces of $^{152}$Gd and its role on GDR.
\end{abstract}

\maketitle
\section{Introduction}
In a simplistic view, the giant dipole resonance (GDR) is due to collective oscillations between protons and neutrons under the influence of the electromagnetic field induced by the emitted/absorbed photons, which results in a large peak in the emission/absorption spectrum of $\gamma$-rays. This fundamental mode of nuclear collective excitation can be built on any state and hence GDR
can probe the structure of the nucleus even at finite temperature ($T$) and angular momentum ($I$) \cite{ALHA,Snov86,Gaar92}.

In a macroscopic theoretical description
of GDR, the GDR observables are coupled with the shape degrees of freedom. Since the nucleus is a tiny system, the thermal fluctuations are more dominant. In the thermal shape fluctuation model (TSFM), the GDR observables are obtained as the average over all the possible shapes of the nucleus \cite{ALHA,Alha89,aruprc1}. In most of such models, the probability of finding the nucleus with a given shape is given in terms of the free energy calculated in a microscopic-macroscopic approach.

In a microscopic way, the GDR can be explained in terms of particle-hole, particle-particle and hole-hole excitations \cite{Gaar92}. One such approach
is the phonon damping model which has been proved to be quite successful in explaining the measured GDR widths ($\Gamma$) as a function of $T$ \cite{DangPLB,Dang:636,Dang:645}.
Recently, these calculations are extended to include the angular momentum for non-collective rotations \cite{DangMAj}. In another microscopic approach, the GDR  in hot and rotating nuclei were studied within the frame work of linear response theory along with static path approximation to the grand canonical partition function \cite{ansari011302,ansari024310}.
 Most of the other microscopic approaches for GDR \cite{Ring,PRL109,PRL111} are not extended to hot and rotating nuclei.

Apart from the theoretical models, phenomenological formulae were also introduced to explain the global properties of $\Gamma$ \cite{Kusn98,Pandit2012434}
in hot and rotating nuclei. These formulae are constructed to fit the experimental $\Gamma$. The results of these phenomenological formulae are similar to those of the TSFM in some cases.

Despite a number of experiments  been carried out to understand the properties of GDR and the effect of $T$ and $I$ on GDR \cite{DRC_NPA,Ishjpg,Ishprc}, still there are several open questions. By large, the measured $\Gamma$ are well interpreted by TSFM calculations, and the phenomenological scaling formula (PSF) \cite{Kusn98} description also was successful in many cases. The failure of PSF and TSFM at low-$T$ is recently analyzed \cite{Pandit2012434} and
the PSF is modified suitably, leading to a critical temperature formula (CTF)
which overcomes the discrepancies at low-$T$. The microscopic origin of such discrepancy is quite known \cite{Dang:531} to be due to pairing and extension of the TSFM to include pairing proved that TSFM is quite successful at low-$T$ also \cite{Rhine_PhysRevC.91,Rhineprc}. Another case where the PSF was observed to be insufficient to explain the measured $\Gamma$ is  the hot and rotating nucleus $^{152}$Gd. It is shown in Ref.~\cite{DRC_JPG} that the PSF could not explain the $\Gamma$ measured at two excitation energies with a single value for the parameter of the PSF (reduced width, $\Gamma_0$). This work highlighted the need for performing proper TSFM calculations specifically for this system to understand the GDR properties.  The CTF was demonstrated  \cite{pandit054327} to be successful and the discrepancy with the PSF was attributed to the role of GDR induced quadrupole moment \cite{Simenel1,Simenel2}.  This argument was generalized to TSFM based on the fact that the PSF mostly mimics the results of TSFM. Simpler TSFM calculations with the liquid drop model
 (LDM) were reported in Ref.~\cite{pandit044325} where the average deformation ($\langle \beta \rangle)$ of $^{152}$Gd (and other selected nuclei) extracted from the measured $\Gamma$
 were shown to be compatible with the results of LDM and CTF.

In this article we present our results from the TSFM calculations and analyze the structure of the hot and rotating $^{152}$Gd. We employ the TSFM built on Nilsson-Strutinsky (NS) calculations with a macroscopic approach to GDR. Our formalism is well tested to explain several GDR observations at higher $T$ and $I$ \cite{aruprc1,aruepja,aruepl,Aruacta}. In a recent work \cite{Ishjpg,Ishprc}, the GDR properties of hot and rotating $^{144}$Sm were studied and the experimental data were very well explained with the TSFM results. A short description about this formalism and our results for the hot and rotating nucleus $^{152}$Gd, are discussed in the forthcoming sections.

\section{Theoretical Framework}
Within the TSFM, the expectation value of an observable $\mathcal{O}$ incorporating both thermal fluctuations and orientation fluctuations is given by~\cite{ALHAO,ALHA} \begin{equation}
\langle \mathcal{O}\rangle _{\beta ,\gamma ,\theta,\phi}=\frac{%
\int_\beta\int_\gamma\int_{\theta}\int_{\phi} \mathcal{D}[\alpha ]\exp \left[{-F_\mathrm{TOT}(T,I;\beta ,\gamma ,\theta,\phi)/T}\right]%
\Im_{\mathrm{TOT}}^{-3/2}\mathcal{O}}{\int_\beta\int_\gamma\int_{\theta}\int_{\phi} \mathcal{D}[\alpha ]\exp\left[{-F_\mathrm{TOT}(T,I;\beta ,\gamma ,\theta,\phi)/T}\right]\Im_{\mathrm{TOT}}^{-3/2}}\;
\label{ave_all}
\end{equation}%
where $\phi,\theta$ are the Euler angles specifying the intrinsic orientation of the nucleus with respect to the axis of rotation. $\beta$ and $\gamma$ are the deformation parameters describing the quadrupole shapes. The volume element is chosen to be $\mathcal{D}[\alpha] = \beta^4
|\sin 3\gamma| \, d\beta \, d\gamma \, \sin\theta \, d\theta \,
d\phi$. $\Im_{\mathrm{TOT}}=\Im_{rig}+\delta\Im$,
where $\Im_{rig}=\Im_{x^\prime
x^\prime}\cos^2\phi\ \sin^2\theta + \Im_{y^\prime
y^\prime}\sin^2\phi\ \sin^2\theta + \Im_{z^\prime z^\prime}
\cos^2\theta$, is the moment of inertia, about an axis with the 
orientation $\theta$ and $\phi$, given in terms of the principal moments of inertia $\Im_{x^\prime x^\prime},\ \Im_{y^\prime y^\prime},\ \Im_{z^\prime
z^\prime}$. $\delta\Im$ is the shell correction for moment of inertia. The total free energy ($F_{\mathrm{TOT}}$) at a fixed deformation is calculated using the Nilsson-Strutinsky  method extended
to high spin and temperature~\cite{aruprc1,aruepja}. 
\begin{equation}
F_{\mathrm{TOT}}=E_{\mathrm{LDM}}+\sum_{p,n}\delta F^{\omega}+\frac{1}{2}\omega({I_\mathrm{TOT}+\sum_{p,n}\delta
I)}\;,  \label{FTOT}
\end{equation}
where $E_{\mathrm{LDM}}$ is the liquid-drop energy and $\delta F^{\omega}=F^{\omega}-\widetilde F^{\omega}$ is the shell correction. $\omega$ and $I_{\mathrm{TOT}}$ are the angular velocity and the total spin, respectively. $\delta I$ is the shell correction to the spin. The microscopic free energy can be calculated using the expression \cite{aruepja}
\begin{equation}
F^\omega= \sum_{i=1}^{\infty}e_i^{\omega}n_i-T\sum_{i=1}^{\infty}s_i\;.
\end{equation}
The single-particle energies ($e_{i}^\omega$) are obtained by diagonalizing the triaxial Nilsson Hamiltonian in a cylindrical representation up to first
twelve major shells. $n_i$ are the occupation numbers given by
\begin{equation}\label{nit}
n_{i}=\frac{1}{1+\exp \left(\frac{e_{i}^{\omega}-\lambda}{T}\right)}
\end{equation}
where $\lambda$ is the chemical potential obtained using the constraint
$\sum_{i=1}^{\infty}n_i=N_p$ and $N_p$ is the total number of particles. $s_i$ are the single-particle entropies and the total
entropy $S=\sum_{i=1}^{\infty}s_i$ can be written as%
\begin{equation}
S=-\sum_{i=1}^{\infty}\left[ n_{i}\ln n_{i}+(1-n_{i})\ln
(1-n_{i})\right]\;.
\end{equation}
$\widetilde F^{\omega}$ is calculated in a Strutinsky way \cite{STRU1} with
exact $T$ and $I$ dependance and more details in this regard can be found in Refs.~\cite{aruprc1,aruepja}. The effect of orientation fluctuations is negligible while calculating the observables like GDR cross sections and the width \cite{Orm20,ansari024310,aruprc1,aruepja}, and hence we have neglected the orientation fluctuations in the present calculations.

The nuclear shapes are related to the GDR observables using a macroscopic
model~\cite{HILT,aruprc1,THIA1} comprising an anisotropic harmonic oscillator potential with a separable dipole-dipole interaction. The Hamiltonian describing GDR excitations can be written as  
\begin{equation}
H=H_{osc}+\eta \ D^{\dagger }D\;.
\end{equation}
Here $H_{osc}$ stands for the anisotropic harmonic oscillator Hamiltonian,
$\eta$ and $D$ represent the dipole-dipole interaction strength and dipole operator, respectively. The total GDR cross section ($\sigma$) is constructed by summing the individual Lorentzians with the peaks at the GDR energies ($E_i$) given by the frequencies corresponding to $H$. The width of these individual components depend on $E_i$ through the relation~\cite{CARL} 
\begin{equation}
\Gamma _{i}\approx 0.026E_{i}^{1.9}\;. \label{PowLaw}
\end{equation}
The GDR full width at half maximum, comparable with the measured
value, is determined from the total GDR cross section, $\sigma$ averaged
over all the possible shapes.

\section{Results and discussion}
\begin{figure}
\includegraphics[width=0.85\columnwidth, clip=true]{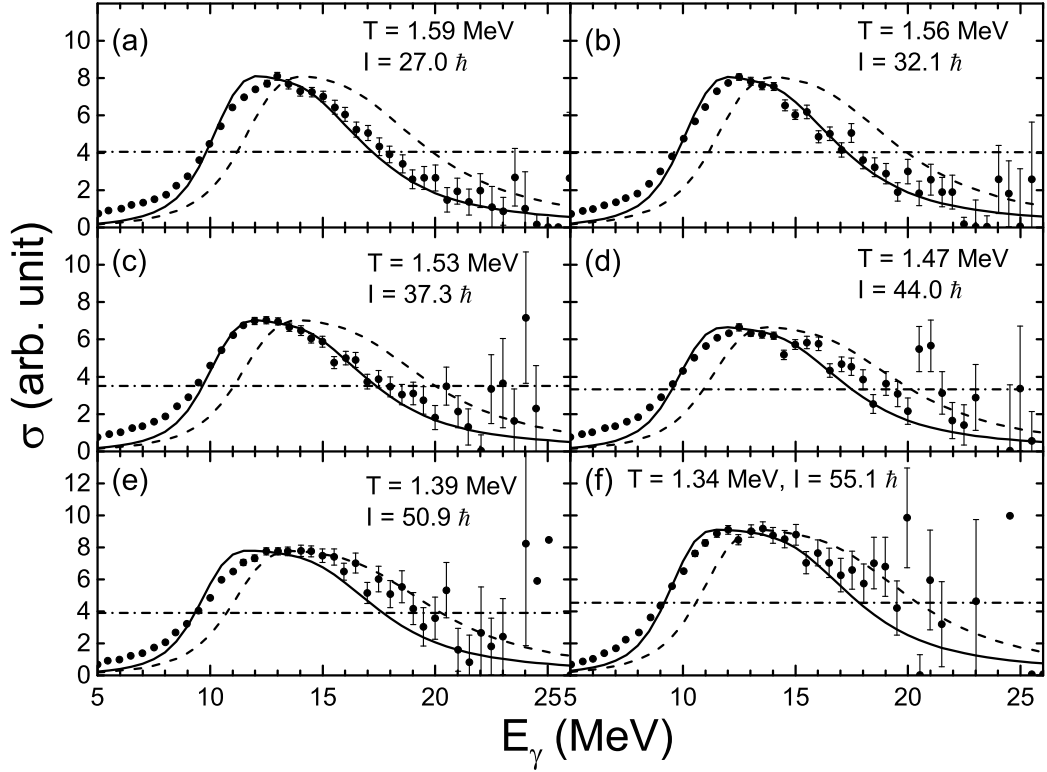}
\caption{The GDR experimental cross sections (filled circles) of $^{152}$Gd at beam energy $E\sim149$ MeV taken from Ref.~\cite{DRC_JPG}, are  compared with the TSFM results obtained using $\eta=2.3$ (solid lines)
and $\eta=3.35$ (dashed lines) at different values of temperature ($T$)
and angular momentum ($I$).}
\label{2006GDR}
\end{figure}

The GDR cross sections and GDR width for the nucleus $^{152}$Gd
are reported in Ref.~\cite{DRC_NPA} for the beam energy, $E\sim149$ MeV and later compiled along with the data at $E\sim185$ MeV in the Ref.~\cite{DRC_JPG}.
TSFM calculations with free energies from LDM were reported in Ref.~\cite{pandit044325}.
We start our analysis by comparing the GDR cross sections calculated with TSFM, and the experimental cross sections at beam energy, $E\sim149$ MeV. These results are presented in Fig.~\ref{2006GDR}. The TSFM calculations are carried out with two different values for the dipole-dipole interaction parameter $\eta$. The value of $\eta$ as $2.3$ is chosen by fitting the experimental cross sections. In this case, we have a good agreement between the experimental and theoretical GDR cross sections represented by the solid lines. However, as shown in Table~\ref{table1}, the corresponding TSFM GDR widths ($\Gamma_{\mathrm{TSFM}}$) are smaller than the experimental GDR widths ($\Gamma_{\mathrm{Expt}}$).
On a careful examination of the cross sections, shown in  Fig.~\ref{2006GDR},
we can note that the experimental data in the higher energy side around the half maximum is quite scattered in most of the cases. Due to this scattered data, the fitting of smooth curves (two component Lorentzian) will carry large errors as reported in Ref.~\cite{DRC_JPG}. These experimental cross sections, with large errors  in the higher energies, tend to yield an overestimated width.

A perfect fit to the GDR cross sections could be possible with a fragmented
GDR spectrum. In a microscopic approach, there can be numerous components
of GDR corresponding to various combinations of particle-hole ($ph$), particle-particle
($pp$), and hole-hole ($hh$) excitations \cite{Speth}. The strength functions for these components can be calculated microscopically but the width is rather artificially introduced. For example, in Ref.~\cite{Dang:645} one can see a fragmented GDR for $^{120}$Sn (Fig.~3 of Ref.~\cite{Dang:645}) with a protruding component around $20$ MeV. Restricting the coupling to $ph$, $pp$, and $hh$ configurations via the doorways can lead to a different scenario (Fig.~5 of Ref.~\cite{Dang:645}). These variations can be understood in terms of the choice of doorway configurations but it is rather difficult to associate a simple physical process. In a macroscopic approach, the picture of splitting of GDR components is quite vivid in terms of the deformation and the Coriolis effects leading to a maximum of five components \cite{HILT}. Even after considering all these effects, our model does not yield such a fragmented GDR consistent with the experimental data. The temperatures corresponding to the observations are sufficiently high so that the macroscopic approach is good enough. The fit achieved for cross sections by our calculations are quite reasonable considering the uncertainties in the data. Hence, the discrepancy in matching the width as shown in Table~\ref{table1}, shall be attributed to the inaccuracy in  experimental cross sections. In other words, with large uncertainties in cross sections, it could be inappropriate to rely on $\Gamma_{\mathrm{Expt}}$ for comparison with the theory. This fact is exemplified by our calculations with $\eta=3.35$ which fits the $\Gamma_{\mathrm{Expt}}$ in a better way as shown in the last column of Table~\ref{table1}. This choice of $\eta$ is far from the systematic values \cite{aruprc1} and yields the cross sections which are very far from the experimental data as shown in Fig.~\ref{2006GDR}. with dashed lines.
  
\begin{table}
\caption{The experimental GDR widths ($\Gamma_{\mathrm{Expt}}$) in $^{152}$Gd at beam energy $E\sim149$ MeV are compared with the  GDR widths calculated
with TSFM ($\Gamma_{\mathrm{TSFM}}$) using two different values for the dipole-dipole
interaction strength parameter $\eta$. The temperature ($T$) and angular
momentum ($I$) correspond to the average values extracted from experimental
data \cite{DRC_JPG}.}
\begin{tabular}{c c c c c}
\hline
\hline
$T$ & $I$ & $\Gamma_{\mathrm{Expt}}$ & \multicolumn{2}{c}{$\Gamma_{\mathrm{TSFM}}$(MeV)} \\
(MeV)& $(\hbar)$ & (MeV )& $\eta=2.3$ & $\eta=3.35$ \\
\hline
\hline
1.59 & 27.0 & 8.5$\pm$0.3 & 7.4 & 8.7 \\ 
1.56 & 32.1 & 8.5$\pm$0.3 & 7.5 & 8.8 \\
1.53 & 37.3 & 8.8$\pm$0.4 & 7.7 & 9.0 \\
1.47 & 44.0 & 8.8$\pm$0.4 & 8.0 & 9.2 \\
1.39 & 50.9 & 8.8$\pm$0.4 & 8.4 & 9.6 \\
1.34 & 55.1 & 10.1$\pm$0.5& 8.7 & 9.8 \\
\hline
\hline
\end{tabular}
\label{table1}
\end{table} 

\begin{figure}
\includegraphics[width=0.85\columnwidth, clip=true]{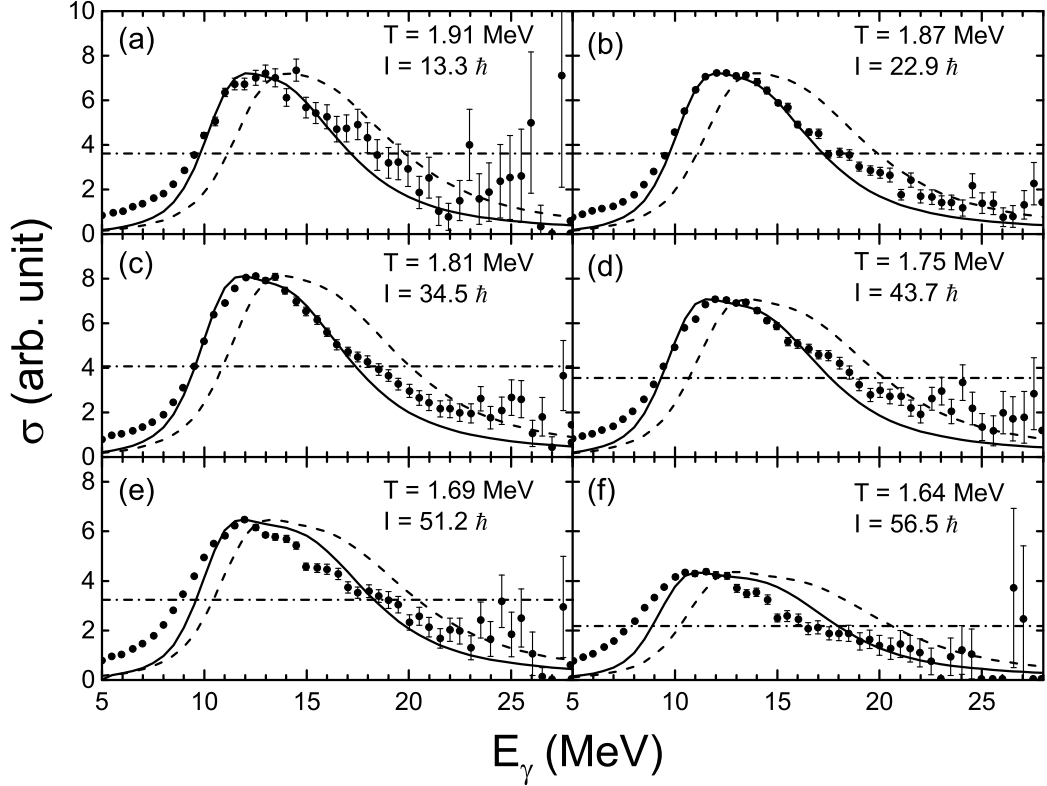}
\caption{Similar to Fig.~\ref{2006GDR} but at beam energy $E\sim185$ MeV.}
\label{2010}
\end{figure}

\begin{table}
\caption{Similar to Table \ref{table1} but at beam energy $E\sim185$ MeV.}
\begin{tabular}{c c c c c}
\hline
\hline 
$T$ & $I$ & $\Gamma_{\mathrm{Expt}}$ & \multicolumn{2}{c}{$\Gamma_{\mathrm{TSFM}}$(MeV)}\\ 
(MeV) & $(\hbar)$ & (MeV) & $\eta=2.3$ & $\eta=3.35$ \\
\hline
\hline
1.91 & 13.3 & 9.5$\pm$0.5 & 7.4 & 8.7 \\  
1.87 & 22.9 & 9.8$\pm$0.3 & 7.5 & 8.8 \\ 
1.81 & 34.5 & 10.0$\pm$0.4 & 7.9 & 9.2 \\  
1.75 & 43.7 & 10.9$\pm$0.5 & 8.3 & 9.6 \\ 
1.69 & 51.2 & 11.1$\pm$0.4 & 8.7 & 9.9 \\ 
1.64 & 56.5 & 11.7$\pm$0.8 & 9.1 & 10.4 \\   
\hline
\hline
\end{tabular}
\label{table2}
\end{table} 

Our results for cross sections at beam energy, $E\sim185$ MeV are presented in Fig.~\ref{2010}, in comparison with the experimental data. In this case also the quality of agreement between our calculations and the data is good except for the very high angular momenta. The corresponding widths are presented in Table \ref{table2}. From these results at  $E\sim185$ MeV, we see the same trend as noticed in the case of $E\sim149$ MeV, viz., (i) the cross sections around the half maximum at the higher energy are scattered leading to overestimation of $\Gamma_{\mathrm{Expt}}$, and (ii) with larger $\eta$, we can obtain a better fit to $\Gamma_{\mathrm{Expt}}$ but the corresponding theoretical cross sections are far from the data. In the present case, the hump at high energy is more pronounced and renders larger width. A simple double Lorentzian fit would suggest a second peak
around 15 MeV which is perhaps inexplicable  within the present theoretical
approach. These facts strengthen our argument that one should not rely on the GDR width, when the cross sections have large uncertainties.  

\begin{figure}[tbp]
\includegraphics[width=.45\columnwidth, clip=true]{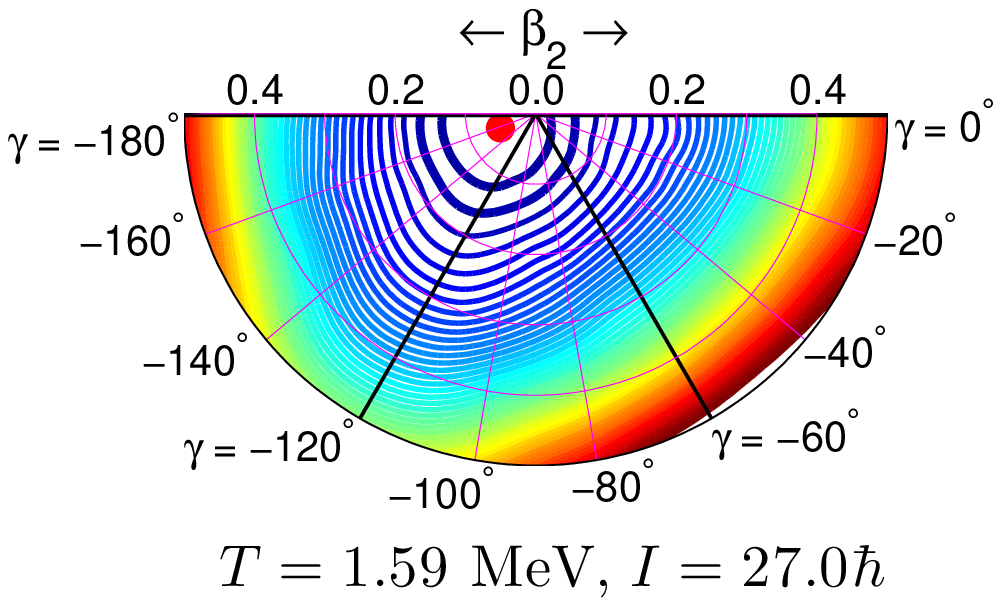}
\includegraphics[width=.45\columnwidth, clip=true]{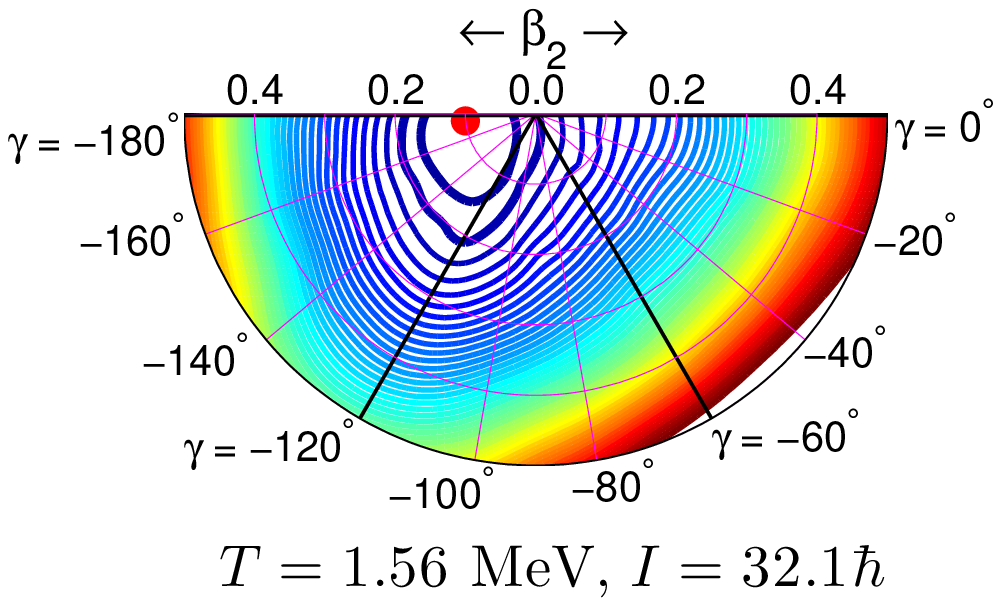}
\vskip-26pt
\hbox{\hspace{30pt}(a) \hspace{200pt}(b)}
\vskip10pt
\includegraphics[width=.45\columnwidth, clip=true]{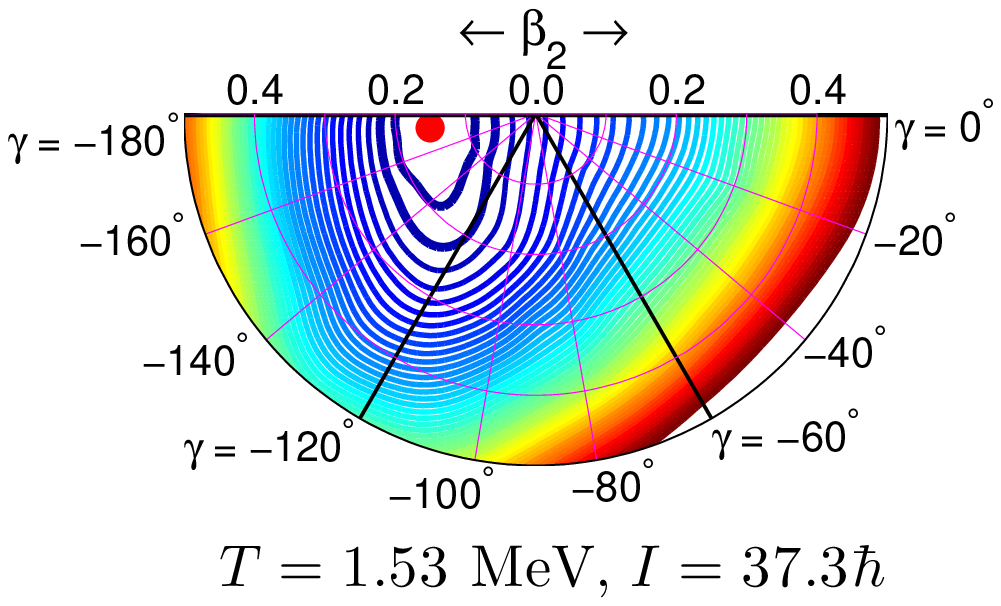}
\includegraphics[width=.45\columnwidth, clip=true]{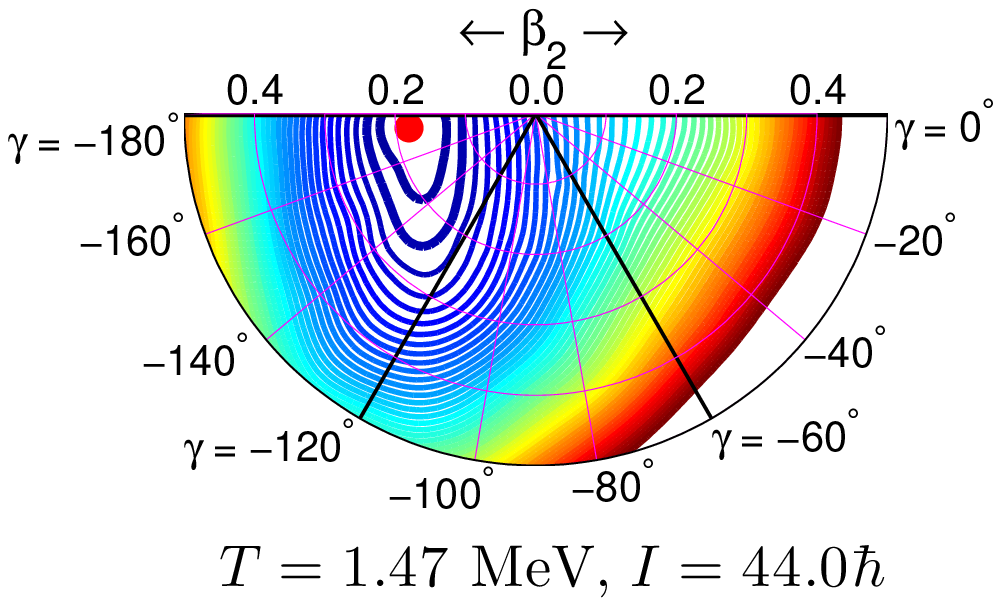}
\vskip-26pt
\hbox{\hspace{30pt}(c) \hspace{200pt}(d)}
\vskip10pt
\includegraphics[width=.45\columnwidth, clip=true]{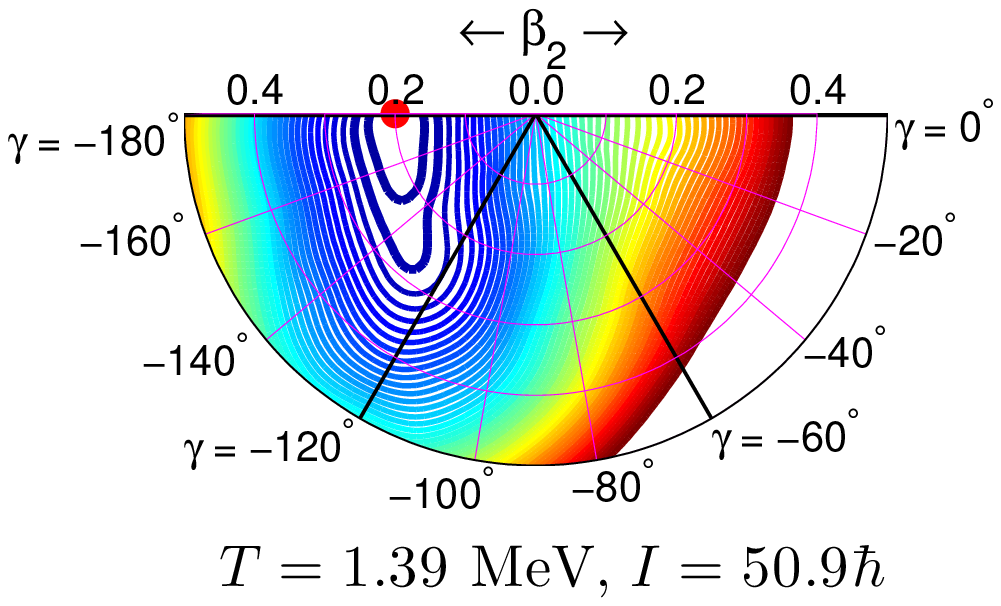}
\includegraphics[width=.45\columnwidth, clip=true]{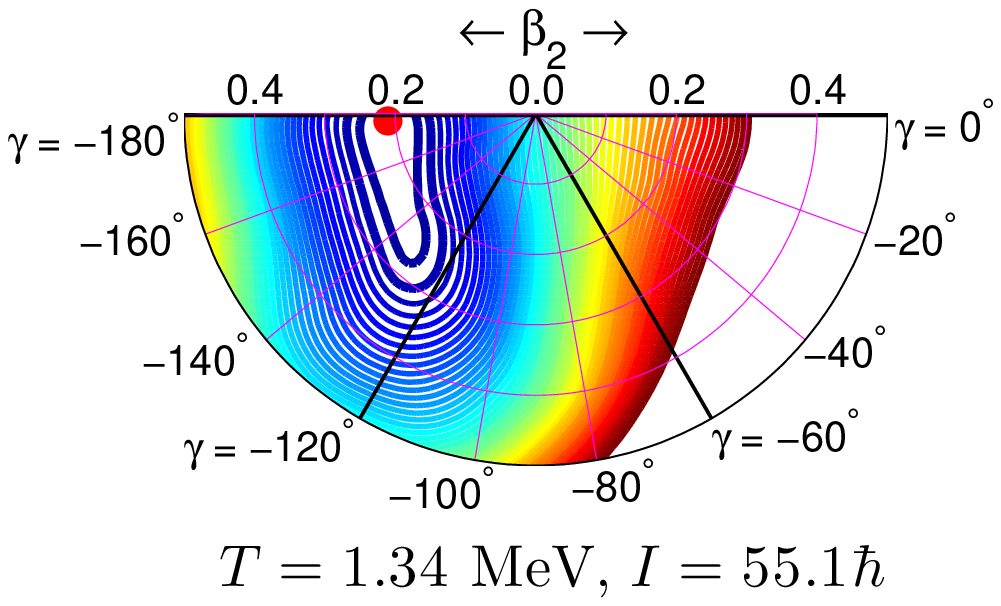}
\vskip-26pt
\hbox{\hspace{30pt}(e) \hspace{200pt}(f)}
\vskip10pt
\caption{(Color online) The free energy surfaces (FES) of $^{152}$Gd at different temperature ($T$) and angular momentum ($I$) combinations corresponding
to the data measured at beam energy $E\sim 149$ MeV.
In this convention, $\gamma=0^\circ$ and $-120^\circ$ represent the non-collective and collective prolate shapes, respectively; $\gamma=-180^\circ$ and $-60^\circ$ represent the non-collective and collective oblate shapes, respectively. The contour line spacing is 0.2 MeV. The most probable shape is represented by a filled circle and first two minima are represented by thick lines.}
\label{2006pes}
\end{figure}

\begin{figure}[tbp]
\includegraphics[width=.45\columnwidth, clip=true]{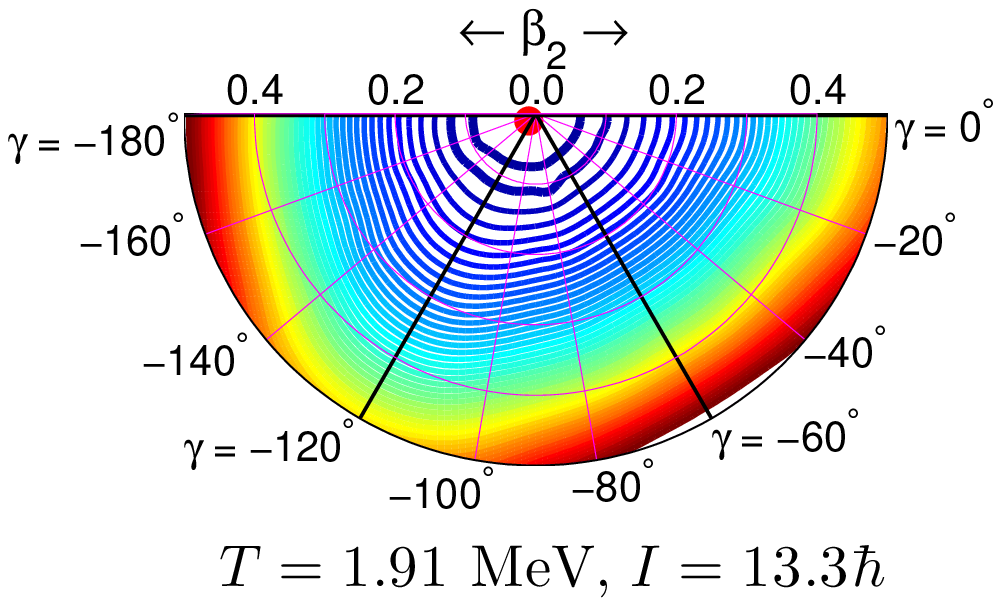}
\includegraphics[width=.45\columnwidth, clip=true]{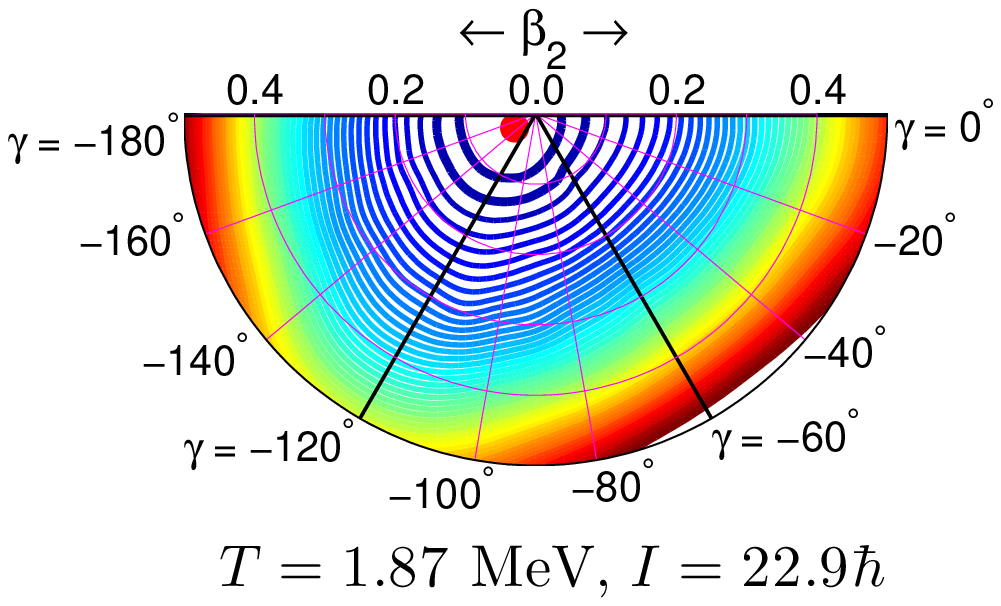}
\vskip-26pt
\hbox{\hspace{30pt}(a) \hspace{200pt}(b)}
\vskip10pt
\includegraphics[width=.45\columnwidth, clip=true]{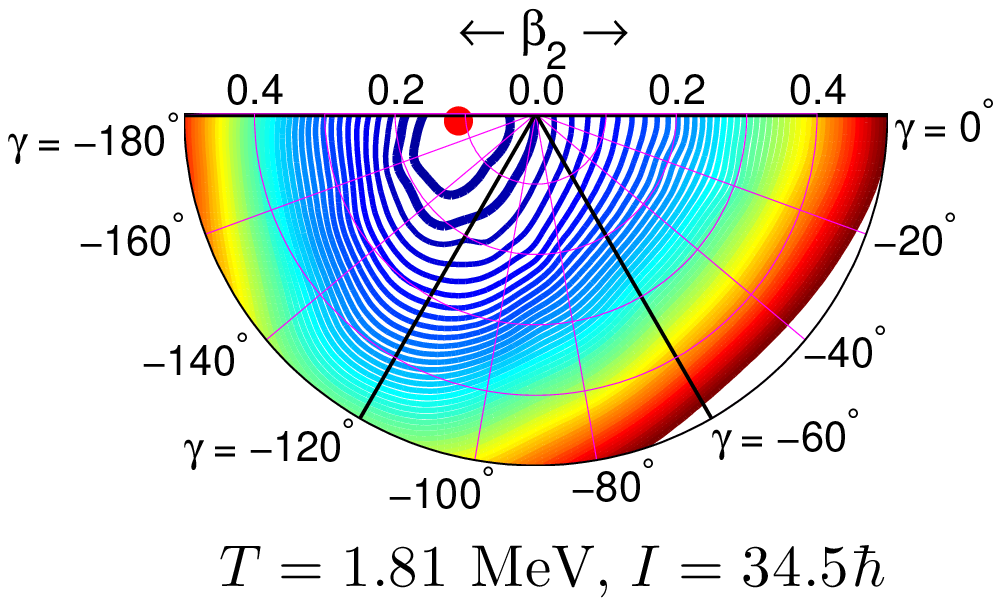}
\includegraphics[width=.45\columnwidth, clip=true]{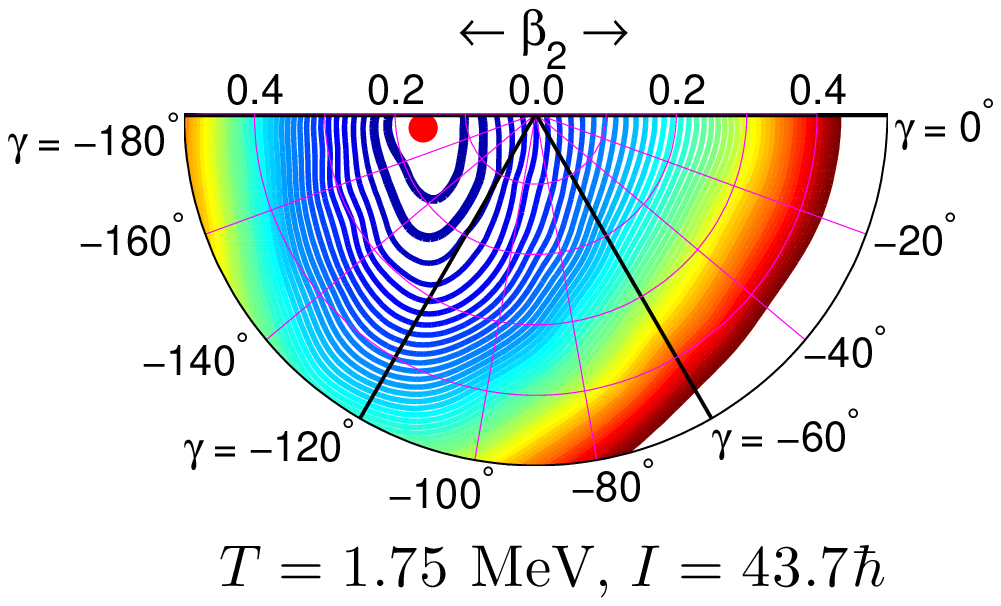}
\vskip-26pt
\hbox{\hspace{30pt}(c) \hspace{200pt}(d)}
\vskip10pt
\includegraphics[width=.45\columnwidth, clip=true]{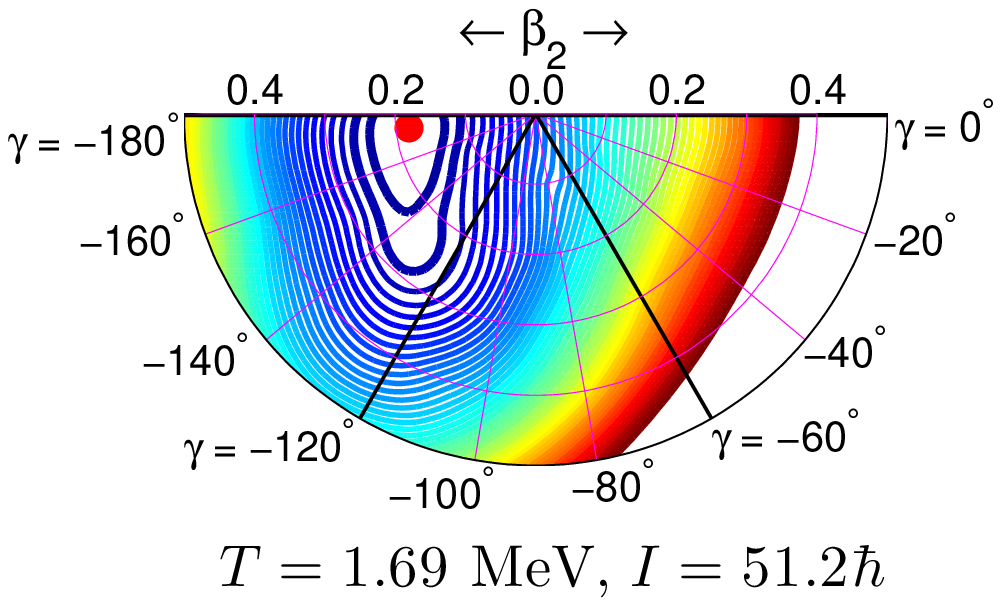}
\includegraphics[width=.45\columnwidth, clip=true]{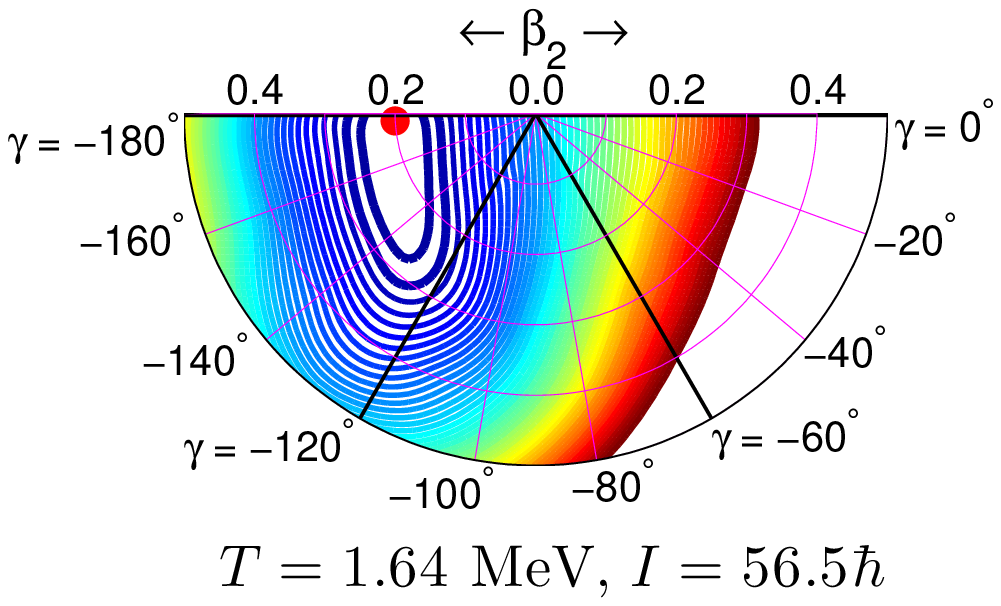}
\vskip-26pt
\hbox{\hspace{30pt}(e) \hspace{200pt}(f)}
\vskip10pt
\caption{(Color online) Similar to Fig.~\ref{2006pes} but for beam energy, $E\sim185$ MeV.}
\label{2010pes}
\end{figure} 
\begin{figure}
\includegraphics[width=0.75\columnwidth, clip=true]{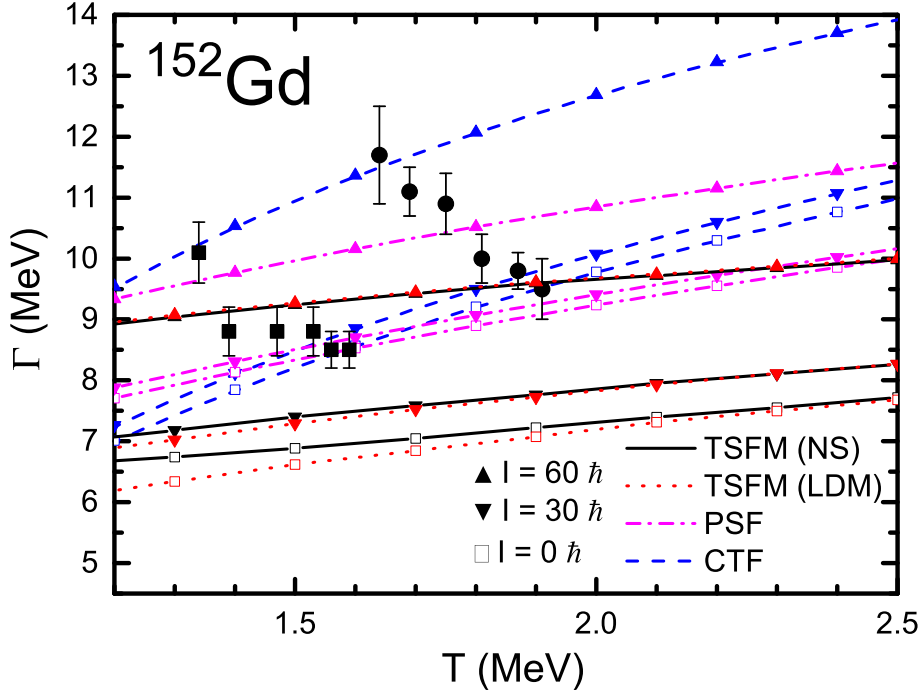}
\caption{(Color online) The GDR width ($\Gamma$) in $^{152}$Gd calculated with various approaches at different angular momenta are plotted as a function of temperature ($T$). The phenomenological scaling formula (PSF) \cite{Kusn98}, critical temperature formula (CTF) \cite{Pandit2012434}, TSFM (LDM), and TSFM (NS) results are represented by dash-dotted, dashed, dotted and solid lines, respectively. The $I=0\hbar, 30\hbar$ and $60\hbar$ are represented by open squares, downward triangles, and upward triangles, respectively. The experimental results are taken for Ref.~\cite{DRC_JPG}, the results at beam energy $E\sim149$ MeV and $185$ MeV are represented by filled squares, and filled circles, respectively.}
\label{PSF_CTF}
\end{figure}

\begin{figure}
\includegraphics[width=0.75\columnwidth, clip=true]{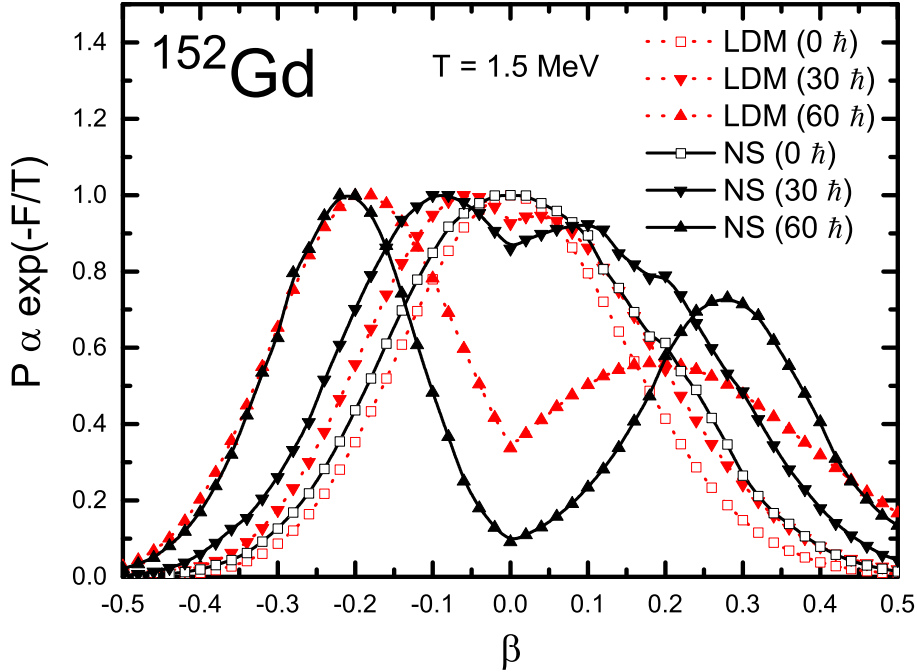}
\caption{(Color online) The probability distribution of $^{152}$Gd shapes calculated with
liquid drop model (LDM) and Nilsson-Strutinsky (NS) approach, at different angular momenta ($I$) are plotted as a function of axial deformation parameter $\beta$. The peaks are normalized to unity.  For the axially deformed shapes, the positive and negative values of $\beta$ can be associated to
$\gamma=-180^\circ$ and $-120^\circ$, respectively. The LDM and NS results are represented by dotted and solid lines, respectively. The results for $I=0\hbar, 30\hbar$ and $60\hbar$ are represented by squares, downward triangles, and upward triangles, respectively.}
\label{Gd_BF}
\end{figure}   

\begin{figure}
\includegraphics[width=0.75\columnwidth, clip=true]{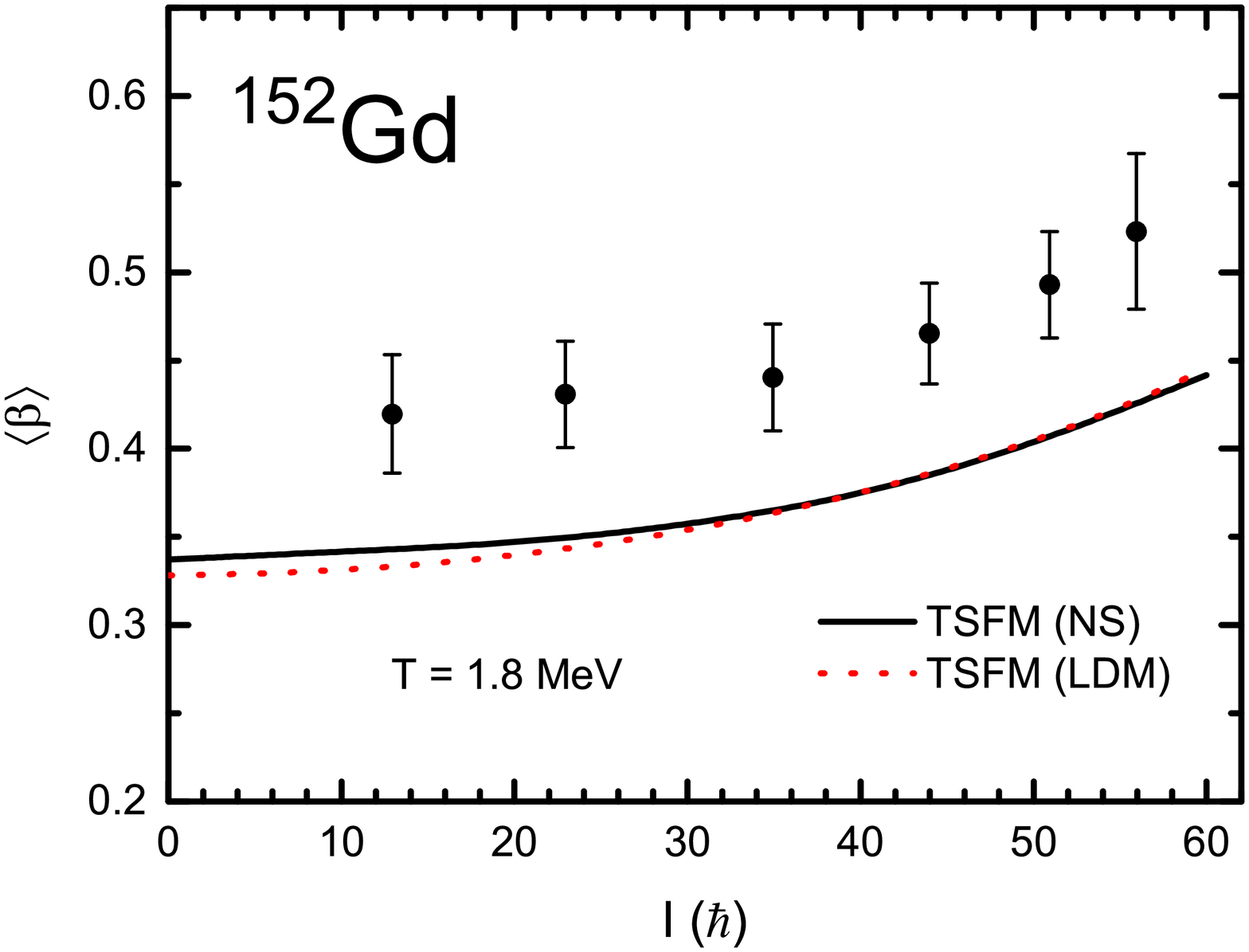}
\caption{(Color online) The average deformation ($\langle \beta \rangle$) of $^{152}$Gd calculated with liquid drop model (LDM) and Nilsson-Strutinsky (NS) approach, at temperature ($T=1.8$ MeV) is plotted as a function of angular momentum ($I$). The LDM and NS results are represented by dotted and solid lines, respectively. The values extracted from GDR data \cite{pandit044325} are represented by filled circles.}
\label{Gd_deepak}
\end{figure}

In order to understand the variation in GDR cross sections in terms of the shape transitions in the nucleus $^{152}$Gd, we present the free  energy surfaces (FES) calculated within our microscopic-macroscopic approach. These
FES at different $T$ and $I$ combinations corresponding to $E\sim149$ MeV
are presented in Fig.~\ref{2006pes}. In general, it is well known that the GDR width ($\Gamma$) is proportional to the average axial deformation parameter $\beta$ ($\langle\beta\rangle$). Within our macroscopic model for GDR, for a given $\beta$, $\Gamma_{\text{Prolate}}> \Gamma_{\text{Triaxial}}>\Gamma_{\text{Oblate}}$. While considering thermal fluctuations, the shallowness of the minimum will lead to pronounced fluctuations and hence larger width. Apart from these shape effects, the GDR width increases with $I$ due to the Coriolis splitting and this change is rapid after $\sim40\hbar$ in this mass region \cite{Kusn98}. From Fig.~\ref{2006pes}, we can note that at high-$T$ and low-$I$, the most probable shape is  oblate with a small deformation. As $I$ increases, the most probable shape changes to a more deformed oblate. At high-$I$, the nucleus shows a clear $\gamma$-softness with the contour for $F\le 0.4$ MeV being quite narrow with respect to $\beta$ ($\sim 0.2$) but spanning the region with $\gamma=-180^\circ$ to $-120^\circ$. In such a case, the GDR samples
a wider variety of shapes and this sampling is enhanced by fluctuations at
finite $T$. For a given $\beta$, since the oblate shape leads to the least
$\Gamma$, the $\gamma$-softness leads to an enhanced $\Gamma$ in comparison
with that of the oblate shape.

Before discussing in detail the role of $\gamma$-softness in $\Gamma$, we
present in Fig.~\ref{2010pes}, the FES corresponding to the $T$ and $I$ combinations extracted from the data measured at $E\sim185$ MeV. In this case, at low-$I$, the most probable shape is spherical; as the $I$ increases the most probable shape changes to an oblate; and at high-$I$, the nucleus shows a clear $\gamma$-softness similar to the previous case. An overall but important observation, from the two sets of FES, is that there are no drastic shape transitions as the $I$ increases. This observation implies that we do not expect any drastic change in the $\Gamma$ with the variation of $T$ and $I$, within the limits suggested by the data at both excitation energies. Secondly, some interesting features seen in the FES need not be reflected in the observables due to the thermal fluctuations which can play a strong role in the considered range of $T$ ($\sim 1.5$ MeV) by smearing out the structural changes caused by shell effects. Such smearing effects can be well depicted while comparing our results with those obtained using the liquid drop model (LDM) FES instead of the FES from the Nilsson-Strutinsky (NS) approach, as shown in Fig.~\ref{PSF_CTF}, where the former and latter results are depicted by dotted and solid lines, respectively. Except for low-$T$ and low-$I$ the TSFM results with FES from LDM and NS, agree well. Thus a LDM description is quite reasonable for the GDR in $^{152}$Gd. Consequently, we observe that some interesting shape effects like the $\gamma$-softness and the shape transitions, as depicted by the FES, are not effectively reflected in the corresponding $\Gamma$. This can be ascribed to the domination of thermal fluctuations which smear out the effects of shape transitions. To see this in detail, we have plotted in Fig.~\ref{Gd_BF} the probabilities corresponding to different shapes as given by the Boltzmann's factor [$\exp(-F/T)$] which can be understood as the weights corresponding to the shapes while averaging the cross sections over different shapes [Eq.~(\ref{ave_all})].

In Fig.~\ref{Gd_BF}, first we would like to draw attention towards the
calculations at $0\hbar$ (open squares) where we can see that the probability
distribution ($P$) is wider in the case of NS calculations and hence yield larger $\Gamma$ (as seen in Fig.~\ref{PSF_CTF}). This difference between LDM and NS results is obviously due to shell effects which would melt as the temperature increases. This difference in $P$ seems to be more at 30$\hbar$
but the contribution from triaxial shapes become significant (not depicted in Fig.~\ref{Gd_BF}) due to the thermal fluctuations and hence the resulting $\Gamma$ are not very different. The dominance of thermal shape fluctuations is quite vivid in the case of $I=60\hbar,$ as evident from the very distinct $P$ for LDM and NS leading to same $\Gamma$. It would be interesting to see how well these shape effects can be reflected by $\Gamma$ at lower $T$ where the shape fluctuations are not that dominant.

The correlation between the shape parameter $\beta$ and $\Gamma$ are well
known \cite{Gall85}. In Ref.~\cite{pandit044325} an empirical correlation is proposed so that one can estimate an ``experimental" deformation ($\beta_{\rm exp}$) from the measured values of $\Gamma$. This allows us to compare the theoretical (average) deformations ($\langle \beta \rangle$) with $\beta_{\rm exp}$ in a way independent of the model for the GDR. Such an analysis has been carried out in Ref.~\cite{pandit044325} for the nuclei $^{59}$Cu, $^{110}$Sn, $^{113}$Sb, $^{152}$Gd, and $^{176}$W. Our results for $\langle \beta \rangle$ in $^{152}$Gd obtained with TSFM calculations are presented in Fig.~\ref{Gd_deepak}.
The TSFM results with LDM and NS methods are represented by dotted and solid lines, respectively and the $\beta_{\rm exp}$ taken from Ref.~\cite{pandit044325} are represented by  filled  circles.
We infer that our TSFM calculations underestimate the deformation and is
explicable as discussed in the case of our results for $\Gamma$ shown in Fig.~\ref{PSF_CTF} and hence strengthen the associated arguments. Here such a discrepancy can be seen as a function of angular momentum ($I$). The $\langle \beta \rangle$ of $^{152}$Gd obtained with the TSFM (LDM) presented in Fig.~3 of Ref.~\cite{pandit044325} shows a higher value, when compared with our results.  Such results based on LDM (at least for $I=0$) should vary smoothly with the mass number but an abrupt raise in $\langle \beta \rangle$ of $^{152}$Gd  could be seen in Fig.~3 of Ref.~\cite{pandit044325}. This raise could be possible with the choice of the parameters in the calculations
which are not known clearly.

Apart from the TSFM results, in Fig.~\ref{PSF_CTF}, we have presented the experimental $\Gamma$ along with the values calculated using the phenomenological
scaling formula (PSF) \cite{Kusn98} and the critical temperature formula (CTF) \cite{Pandit2012434}. In the PSF, the ground state GDR width is assumed
to be $\Gamma_0=3.8$ MeV all the other parameters are fitted empirically in a global way. As shown in Ref.~\cite{pandit054327}, the CTF can explain the enhanced $\Gamma$ whereas the PSF fails in this regard. However, a larger $\Gamma_0$ of 5.7 MeV is used to reproduce the measured $\Gamma$. Another parameter of CTF is the critical temperature, $T_c$, which is also adjusted to get a better fit with the experimental results. Naturally, with the help of two adjustable parameters, the CTF could explain the data much better than the PSF. These parameters can change the rate at which $\Gamma$ changes with $I$ and $T$.  At $T=1.2$ MeV, the difference in $\Gamma$ values  with PSF (dash-dotted lines) at two extreme values of $I$ is, $\delta \Gamma \sim1.6$ MeV. Whereas, with CTF (dashed lines), $\delta \Gamma \sim 2.5$ MeV. So the CTF parameters are chosen to have a stiff $\Gamma$ with respect to both $I$ and $T$, and hence the CTF is able to explain the $\Gamma_{\mathrm{Expt}}$ in a larger range. Results from such a stiff parameterization
cover more regions in Fig.~\ref{PSF_CTF} and eventually the CTF encompasses all the experimental data. The success of CTF in explaining the $\Gamma$
in an empirical manner shall not be considered as a validation of the extracted
$\Gamma$, as the reliable information lies within the GDR cross sections.

\section{Summary}
The thermal shape fluctuation model (TSFM) study, of the  giant dipole resonance (GDR) in the hot and rotating nucleus $^{152}$Gd, reveals that there are no major anomalies in the values measured at two excitation energies. The GDR cross sections calculated with standard parameters are in agreement with the experiment, except for a shoulder around 17 MeV where the uncertainties are large. The component of GDR leading to such a shoulder cannot be explained
within a macroscopic approach for GDR where the splitting due to the deformation
and Coriolis effects are properly taken care. It could be instructive to
see whether such a high energy component can be explained with fully microscopic
approaches for GDR. It will also be useful to have a careful introspection of the data to ascertain the shoulder around 17 MeV. Due to this shoulder, the corresponding GDR widths are overestimated and hence their comparison with calculated values does not convey the complete information.
Tuning the phenomenological parameterizations to reproduce such widths
conceal the interesting information contained in the cross sections. The free energy surfaces in $^{152}$Gd calculated at higher angular momenta show a clear gamma softness which could have led to larger GDR widths. However, such shape aspects are smeared by the thermal fluctuations which are dominant in the considered temperatures. It would be interesting to see how such shape effects survive at lower temperatures.

\section*{Acknowledgments}
The authors acknowledge the financial support from  the Science and Engineering Research Board (India), SR/FTP/PS-086/2011 and DST/INT/POL/P-09/2014. The financial support from the Ministry of Human Resource Development, Government of India, to one of the authors (A.K.R.K) is gratefully acknowledged.

\end{document}